\documentclass[onecolumn,draftcls]{IEEEtran}

\usepackage{times} 
\usepackage{graphicx}                              
\usepackage{amssymb}
\usepackage[cmex10]{amsmath}
\usepackage{subfig}
\usepackage{color}
\definecolor{MyGray}{rgb}{0.95,0.95,0.95}

\newcommand{\reffig}[1]{Figure~\ref{#1}}
\newcommand{\reftab}[1]{Table~\ref{#1}}

\newcommand*{\textfrac}[2]{\ensuremath{\kern.1em
\raise.5ex\hbox{\the\scriptfont0 #1}\kern-.1em
/\kern-.15em\lower.25ex\hbox{\the\scriptfont0 #2}}}

\usepackage{cite}

\ifCLASSINFOpdf
\else
\fi
\begin{document}
%
\title{On Complexity, Energy- and Implementation- Efficiency of Channel Decoders}
%
%
%

\author{Frank~Kienle,~\IEEEmembership{Member,~IEEE,}~ Norbert~Wehn,~\IEEEmembership{Senior Member,~IEEE},
        and~ Heinrich~Meyr,~\IEEEmembership{ Fellow,~IEEE}
\thanks{{This work has been partly supported by the UMIC Research Center, RWTH
Aachen University. F. Kienle and N. Wehn are with the Microelectronic Systems Design Research Group, University of Kaiserslautern, Kaiserslautern, Germany, e-mail: \{kienle,wehn\} @eit.uni-kl.de}
{H. Meyr is with the Institute for Integrated Signal Processing System, RWTH Aachen University, Aachen, Germany}}}
%
%

\markboth{IEEE Transactions on Communications,~Vol.~x, No.~x, xxxx}%
{Shell \MakeLowercase{\textit{et al.}}: Bare Demo of IEEEtran.cls for Journals}

%



\maketitle

\begin{abstract}

Future wireless communication systems require efficient and flexible baseband receivers.
Meaningful efficiency metrics are key for design space exploration to quantify the algorithmic and the implementation complexity of a receiver. Most of the current established efficiency metrics are based on counting operations, thus neglecting important issues like data and storage complexity.

In this paper we introduce suitable energy and area efficiency metrics which resolve the afore-mentioned disadvantages. These are decoded information bit per energy and throughput per area unit. Efficiency metrics are assessed by various implementations of turbo decoders, LDPC decoders and convolutional decoders. 
New exploration methodologies are presented, which permit an appropriate benchmarking of implementation efficiency, communications performance, and flexibility trade-offs. These exploration methodologies are based on efficiency trajectories rather than a single snapshot metric as done in state-of-the-art approaches.

\end{abstract}


%
\IEEEpeerreviewmaketitle

\section{Motivation}


Today, high-end smart phones have to support multiple radio standards, advanced graphic-
and media applications and many other applications resulting in a workload of about 
100 giga operations per second in a power budget of 1 Watt \cite{ber_09}. 
The baseband processing in the radio part (mainly front-end processing, demodulation and 
decoding) requires more than 50\% of the overall workload in a state-of-the-art 3.5G smart 
phone. To achieve higher spectral efficiency new transmission techniques like MIMO will be established. 
However, this will increase the workload even further. Thus there is a strong 
need for {\em efficient} wireless baseband receivers. The overall efficiency of a baseband 
receiver depends on

\begin{itemize}
\item {\em communications efficiency}: expressed by the spectral efficiency and signal-to-noise ratio (SNR). The requirements on the communications efficiency 
have the largest impact on the selected baseband processing algorithms. 
\item  {\em implementation efficiency}: related to silicon area, power and energy.
Here, the energy efficiency is the biggest challenge due to the limited available battery 
power in many devices.   
\item  {\em flexibility}: in software defined radio, receivers have to support multiple standards and should be 
configurable at run-time (see software defined radio). There are various silicon 
implementation styles ranging from
general purpose architectures, over DSPs and ASIPs down to fully physically optimized IP blocks
which strongly differ in their implementation efficiency but also in their flexibility. 
For each building block of the receiver a detailed 
analysis of flexibility requirements has to be carried out to find the best 
flexibility/cost trade-off. Thus, advanced baseband receivers are heterogeneous 
multi-core architectures implemented in different design styles.  
\end{itemize}

System requirements are very often specified by communication standards like UMTS, 
LTE and WiMAX, which define different 
services in terms of required communications performance and system data throughput, i.e. 
information bits per second.
To obtain an efficient baseband implementation, a careful and
elaborate \emph{design space exploration} has to be performed. This is a very challenging 
task due to the size and the multi-dimensionality of the space.
Therefore it is mandatory to prune the design space in an early stage of the design 
process. In this process the algorithms have to be selected and quantitatively compared 
to each other with respect to their system performance and implementation efficiencies. 

Appropriate metrics are key for efficient design space exploration to measure the 
algorithmic and the implementation complexity respectively.

\subsection{Algorithmic Complexity}

\begin{figure}[t]
\centering
\includegraphics[width=1.0\columnwidth]{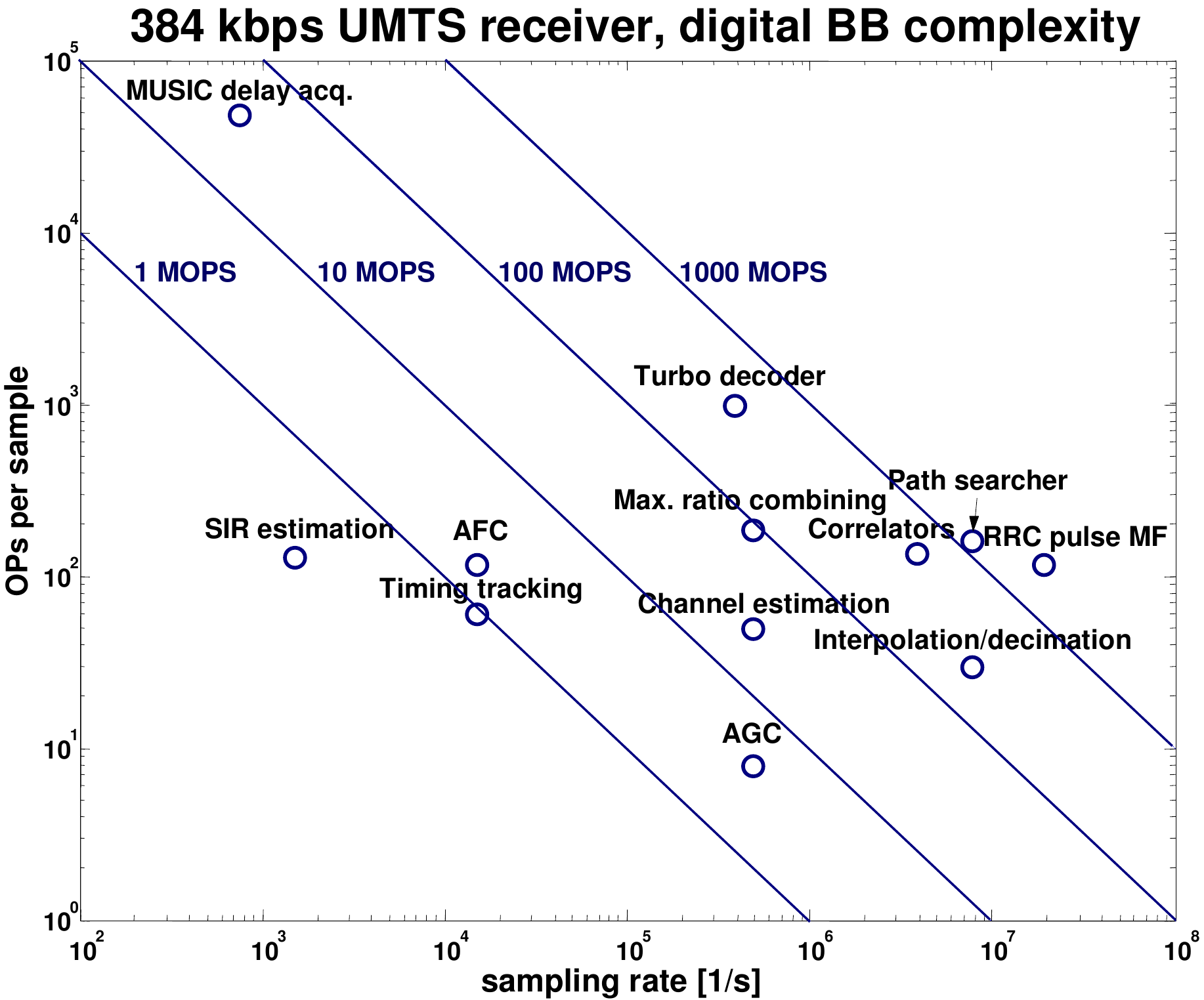}
\caption{Operations per sample over sampling rate [1/s]}
\label{fig:sampling_ops}
\end{figure}

There exists no universal measure for complexity. In computer science the O-notation describes the asymptotic complexity behavior of an algorithm. In information theory the Kolmogoroff complexity is defined by the minimum description length of a string. These measures are inadequate for implementation purposes. A useful description of complexity for our purpose is to use the number of "algorithmic" operations which have to be performed per received samples by the algorithms of a baseband receiver. This complexity metric has the advantage of being independent of a specific implementation of the algorithms.
Based on this complexity definition a two-dimensional graph can be set up in which the horizontal axis correspond to the sample rate of the receiver (which is proportional to the data rate) and the vertical axis corresponds to the operations per sample which have to be carried out. Typically, both axes are scaled logarithmically. 
An example of such a graph is show exemplarily in \reffig{fig:sampling_ops} for the digital baseband processing of a 384 kbit/s UMTS receiver.
The off-diagonal lines describe points of equal number of \emph{operations per second}, often expressed in million of operations per second (\textit{MOPs})or giga operations per second (\textit{GOPs}). 

Several important conclusions can be draw form this figure. First, the receiver task is heterogeneous. There exist a large variety of algorithms ranging from the complex MUSIC algorithm to the simple root raised cosine matched filtering. Note, that the largest number of operations is performed in simple algorithms such as filtering and correlation while the most complex MUSIC algorithm in this example requires less \textit{MOPs}. From this follows that counting only operations/sec is entirely misleading. The second conclusion is that the heterogeneity of the algorithms points to architectural features for implementation. Simple algorithms such as filtering and correlation can be implemented very efficiently in architectures requiring little flexibility in the form of  parameterizability. 
Complex algorithms must be programmable and thus require high flexibility.
 Recently, Van Berkel determined the complexity of various algorithms in baseband processing based on a similar metric. In his remarkable and comprehensive \cite{ber_09} he has shown the number of "algorithmic" operations which have to be performed per received bit by the algorithms of a baseband receiver for different communication standards.  
Eberli from ETH Z\"urich 
is using a similar metric \cite{PhdEberli09} for measuring complexity in baseband
processing by calling these operations "atomic" operations. 



From a communication system point of view we can separate digital processing in the baseband into two
parts: the so called "inner modem" and "outer modem" \cite{meymoe_98} respectively. Task of the inner modem 
is the extraction of symbols from the received signal waveform, i.e., equalization, channel estimation, 
interference cancellation and synchronization. The outer modem performs demodulation, de-interleaving 
and channel decoding on the received symbols. Thus algorithmic complexity in baseband processing is 
normally separately plotted for the inner and the outer modem respectively. A large diversity exists in the 
various baseband processing algorithms with respect to operation types, operation complexity, and data types
especially between the inner and out modem.
Figure 2 in \cite{ber_09} shows the algorithmic complexity for the inner and outer modem measured in
giga operations per second (GOPs). It can be seen that sophisticated decoding schemes like turbo and LDPC codes
utilized in advanced services like LTE require much more operations than the 
algorithms of the inner modem.

\subsection{Implementation Complexity}

On the implementation side a strong emphasis has to be put on the energy efficiency. Implementation complexity and algorithmic complexity are strongly interrelated in wireless
baseband processing. Thus they have to be related to each other. Often it is argued that the 
implementation complexity is directly related to the algorithmic complexity. E.g.
Eberli \cite{PhdEberli09} considers the implementation complexity by introducing 
a cost factor for each atomic operation which reflects its implementation cost. 
For design space exploration, graph representations are commonly used:

\begin{itemize}
\item A two dimensional energy efficiency graph: one axis corresponds to the algorithmic complexity, 
e.g. measured in $GOPs$, and the other axis to the power, e.g. measured in $mW$, consumed when providing 
the corresponding operations/second.
Each point in this graph describes the {\em energy efficiency metric}, i.e. {\em operations/second/power unit}, 
usually measured in $GOPs/mW$. Since energy corresponds to power multiplied with execution time, each point 
gives the {\em operations/energy} measured in $operations/Joule$.
\item In a similar way we can set up an area efficiency graph in which one axis represents the needed area.
Each point in this graph yields the {\em area efficiency metric}, i.e.  {\em operations/second/area unit}, 
usually measured in $GOPs/mm^2$. 
\end{itemize}

Note that the energy and area efficiency for the same
algorithmic complexity can vary by several orders of magnitude,
dependent on the selected implementation style. By far the highest energy efficiency is achieved by physically optimized circuits, however, at the expense of no flexibility. The highest flexibility via software programmability at the expense of low energy efficiency is achieved by digital signal processors. The designer has to find a compromise between the two conflicting goals by trading off flexibility vs. energy efficiency. Flexibility is hard to quantify. The optimum design point is thus to be understood qualitatively. It depends on the application and a large number of economic and technical considerations. We can combine energy and area efficiency in a two dimensional
design space in which the two axis correspond to
area and energy efficiency respectively. This is a well known
representation of the design space.

%

\subsection{Assessment of Metrics}

Area, throughput and especially energy in many system-on-chip implementations are dominated by 
data-transfers and storage schemes \cite{mirghe_02} and not by the
computations itself. However common metrics as described above are focusing solely  on operations, 
and are not considering data-transfer and storage issues at all. Thus, these metrics are only valid
if the operations dominate the implementation complexity. This is the case in data-flow dominated algorithms like
an FFT calculation, correlation or filtering. 

Most algorithms in the inner modem of baseband receivers belong to this
class of algorithm. However the \textit{channel decoding algorithms} in the outer modem largely differ 
from the algorithms used in the inner modem. Here, the operations to be performed are non-standard
operations (e.g. tanh) using non-standard data types (e.g. 7 bits). But more important, the overall implementation complexity, 
especially energy, is dominated by data-transfers and storage schemes. A change in the algorithm with 
respect to computation, e.g. optimal versus suboptimal algorithm by approximating computations, 
has only a minor impact on the energy efficiency as shown later.

The transitions from 3G to LTE advanced require 2 orders of magnitude improvement in energy efficiency. This improvement will come to a small extent from technology scaling \cite{row_09}. Efficient system-on-chip implementations are feasible when channel coding schemes and
the corresponding  decoding algorithms are co-designed together with
the architecture (architecture aware code design) \cite{boucas_00}\cite{mansha_03}\cite{kwalee_02}. This is in accordance with a general trend towards  co-design algorithm and architecture in receiver design realizing that the traditional separation of algorithm and architecture design leads to  suboptimal results.  In channel decoding the co-design focuses on data-transfer and storage schemes. Examples are special
interleavers for turbo codes (e.g. LTE standard \cite{3gpp}) and special
structures of the parity check matrix for LDPC codes (e.g.
DVB-S2 standard \cite{eudigital}). These special structures allow an efficient
parallel implementation of the decoding algorithm with small
overhead in data-transfer and storage. GOPs based metrics
do not at all reflect such specific structures.


Another important issue is flexibility. Flexibility on the algorithmic side, e.g., code rates 
and block sizes in the case of channel decoding, have a large impact on the implementation
complexity. By looking only on the operations in the algorithm, flexibility is normally not considered.

In summary, efficiency metrics based on GOPs are questionable. Particularly, for non-data flow dominated algorithms
since they entirely neglect important issues like data and storage complexity, algorithm/architecture 
co-design and flexibility.

In this paper we focus on channel decoding as application. The contributions of this paper
are:

\begin{itemize}
	\item we will show that the GOPs metric yields wrong conclusions.
	\item we will introduce suitable metrics for energy and area efficiency.
	\item we present a methodology for design space exploration based on these metrics. 
\end{itemize}

\begin{table*}[th!]
\begin{center}
\begin{tabular}{|p{0.8in}|p{0.8in}|p{0.4in}|p{1.0 in}|p{0.5in}|p{0.4in}|p{0.4in}|}
\hline
\textbf{Decoder} & \textbf{Flexibility} & \textbf{Max. Block-size} & \textbf{Throughput [Mbit/s]} & \textbf{Frequency [MHz]} & \textbf{Area [mm2]} & \textbf{Dynamic Power [mWatt]} \\ 
\hline
ASIP \cite{vogweh_08}  & Conv. Codes\newline Binary TC\newline Duo-binary TC & N=16k & 40 \newline 14 (6iter) \newline 28 (6iter) & 385 (P\&R) & 0.7 (P\&R) & \~{}100 \\ 
\hline
LTE turbo \cite{mayiln_10} & R=1/3 to R=9/10 \newline by puncturing & N=18k & 150  (6.5 iter) & 300 (P\&R) & 2.1 (P\&R) & \~{}300 \\ 
\hline
LDPC flexible & R=1/4 to R=9/10 & N=16k & 30  (R=1/3 40iter) \newline 100  (R=1/2 20iter) \newline 300  (R=0.83 10iter)  & 385 (P\&R) & 1.172 (P\&R) & \~{}389 \\ 
\hline
LDPC WiMedia 1.5 \cite{allber_09} & R=1/2-4/5  & N=1.3k & 640 (R=1/2 5iter) \newline 960 (R=0.75 5iter) & 265 & 0.51 & \~{}193 \\ 
\hline
CC Decoder & 64-state NSC & ~ & 500 & 500 & 0.1 & \~{}37 \\ \hline
\end{tabular}
\end{center}
\caption{Reference decoders: service parameters and implementation results in 65nm technology} 
\label{tab:IMPL}
\end{table*}

\section{Reference Designs}

Reference designs are key to assess various metrics. Thus,
we selected 5 different channel decoder implementations which our research group has designed
in the last couple of years. Using own design has the advantage that all data are available.
The decoders differ in services (throughput, block sizes, code rates), 
decoding algorithms, flexibility and implementation 
styles. Selected codes are convolutional codes, turbo codes and LDPC codes. 
The 5 different decoders are:
\begin{itemize}
 \item An application specific instruction set processor (ASIP) \cite{vogweh_08} capable of processing 
 binary turbo codes, duo-binary turbo codes and various convolutional codes with different throughputs
 dependent on code rate and decoding scheme.
 \item A turbo decoder which is LTE \cite{3gpp} compliant. The maximum throughput is 150Mbit/s at 6.5 
 decoding iterations.
 \item An LDPC decoder optimized for flexibility, supporting two different decoding algorithms, code rates 
 from R=1/4-9/10 and a maximum block length of 16384.
 \item An LDPC decoder which is WiMedia 1.5 compliant and optimized for throughput, supporting code rates from R=1/2-4/5 
 with two block lengths N=1200 and N=1320 bits \cite{allber_09}.
 \item A convolutional decoder with 64-state which is WiFi \cite{iewireless} compliant.
 \end{itemize}

All decoders are synthesized on a 65nm CMOS technology under worst case conditions with $V_{dd}=1.0V, 120 °C$.
Power estimations are based on nominal case $V_{dd}=1.1V$. \reftab{tab:IMPL} gives an overview of the key parameters 
of the different decoders. P\&R indicates that the corresponding data are post-layout data. The payload (information bits) throughput
depends on the number of decoding iterations for turbo and LDPC codes which also impacts the communications 
performance. Thus, the throughput is specified dependent on the number of iterations.

In \reftab{tab:GOPS} we show the number of algorithmic operations required to process the different types of convolutional codes, turbo codes and LDPC codes. Bit-true C-reference models are used for operation counting. All operations were normalized
to an 8 bit addition. The number of operations is related to one information bit which has to be decoded, 
i.e., operations/information bit.
The total number of operations which have to be performed per second, i.e. GOPs, depends on the code rate R and throughput
which depends on the number of iterations for LDPC and turbo codes. Two different algorithms for LDPC codes were investigated. 
Both algorithms are suboptimal
algorithm approximating the belief propagation algorithm: the Min-Sum algorithm with a
scaling factor and the $\lambda$-3-Min algorithm \cite{guibou_03}
which is a more accurate approximation.  However the latter one needs
about 3.3 times more operations. This more accurate approximation is mandatory if 
lower code rates $R<0.5$ have to be supported like in DVB-S2 decoders \cite{mulsch_09}.
In \reftab{tab:GOPS} we have only listed the operations for the Min-Sum algorithm. To obtain the operations
for the $\lambda$-3-Min algorithm, operations and GOPs have to be multiplied by 3.3 respectively. The flexible 
LDPC decoder in our reference design was designed for both decoding algorithms, the WiMedia LDPC 
decoder is based on the Min-Sum algorithm only.

\begin{table}[t]
\begin{center}
\begin{tabular}{|p{0.55in}|p{0.3in}|p{0.35in}|p{0.4in}|p{0.4in}|p{0.4in}|}
\hline
\textbf{Code} & \multicolumn{2}{|p{0.7in}|}{\textbf{operations/bit} } & \multicolumn{3}{|p{1.2in}|}{\textbf{GOPs (w.r.t. throughput)}} \\ 
\hline
 			 		& \#iter & \#(op/bit) & 100 Mbit/s 	& 300 Mbit/s 	& 1 Gbit/s \\
\hline
CC (states=64)     	 &    	&{200} 					& 20   		&  60 		& 200  \\
\hline
    			   	& 5 iter 	& 75/R 			& 7.5/R  	& 22.5/R 	&  75/R  \\ 
\cline{2-6}
LDPC 			 	& 10 iter 	& 150/R 		& 15/R   	&  45/R 	&  150/R \\ 
\cline{2-6}
Min-Sum 			& 20 iter 	& 300/R 		&  30/R  	&  90/R	 	&  300/R \\ 
\cline{2-6}
 			  	& 40 iter 	& 600/R 		&  60/R  	&  180/R 	&  600/R \\ 
\hline
Turbo 			  	& 2 iter 	& 280 			&  28  		&  84 		&  280  \\ 
\cline{2-6}
(Max-Log) 		  	& 4 iter 	& 560 			&  56 		& 168 		&  560  \\ 
\cline{2-6}
 \multicolumn{1}{|c|}{}    	& 6 iter 	& 840 			&  84 		& 252 		&  840  \\ 
\hline
\end{tabular}
\end{center}
\caption{Number of normalized algorithmic operations per decoded information bit for different channel decoders dependent
on throughput and code rate R.}
\label{tab:GOPS}
\end{table}

\section{Suitable Metrics}

\begin{figure}[t]
\centering
\includegraphics[width=1.0\columnwidth]{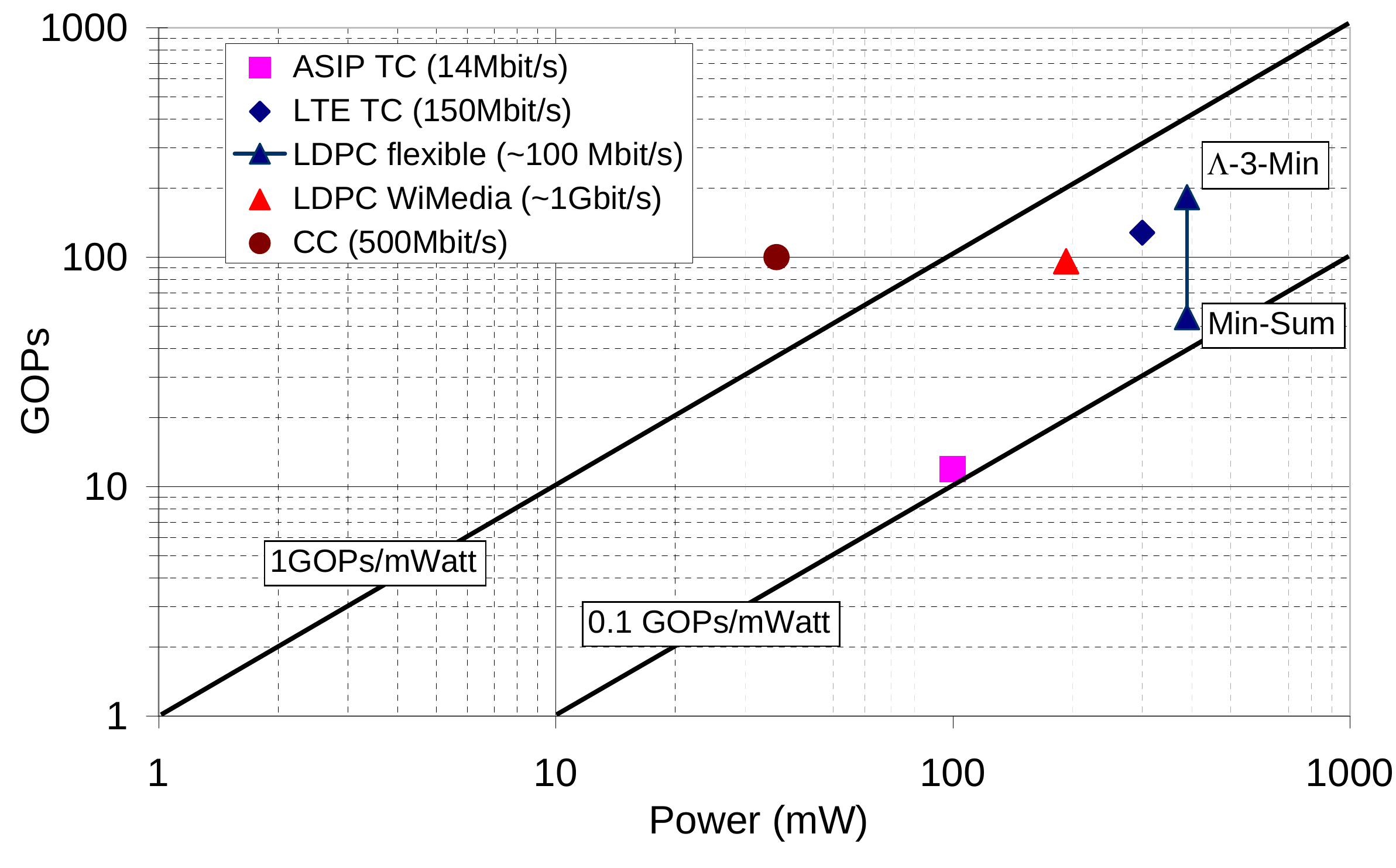}
\caption{Operations/second versus power}
\label{fig:gops_power}
\end{figure}

The energy efficiency graph for our reference designs is shown in \reffig{fig:gops_power}. It can
be seen that the energy efficiency, measured in GOPs/mW, largely varies for the
different decoders.

\begin{figure}[t]
\centering
\includegraphics[width=1.0\columnwidth]{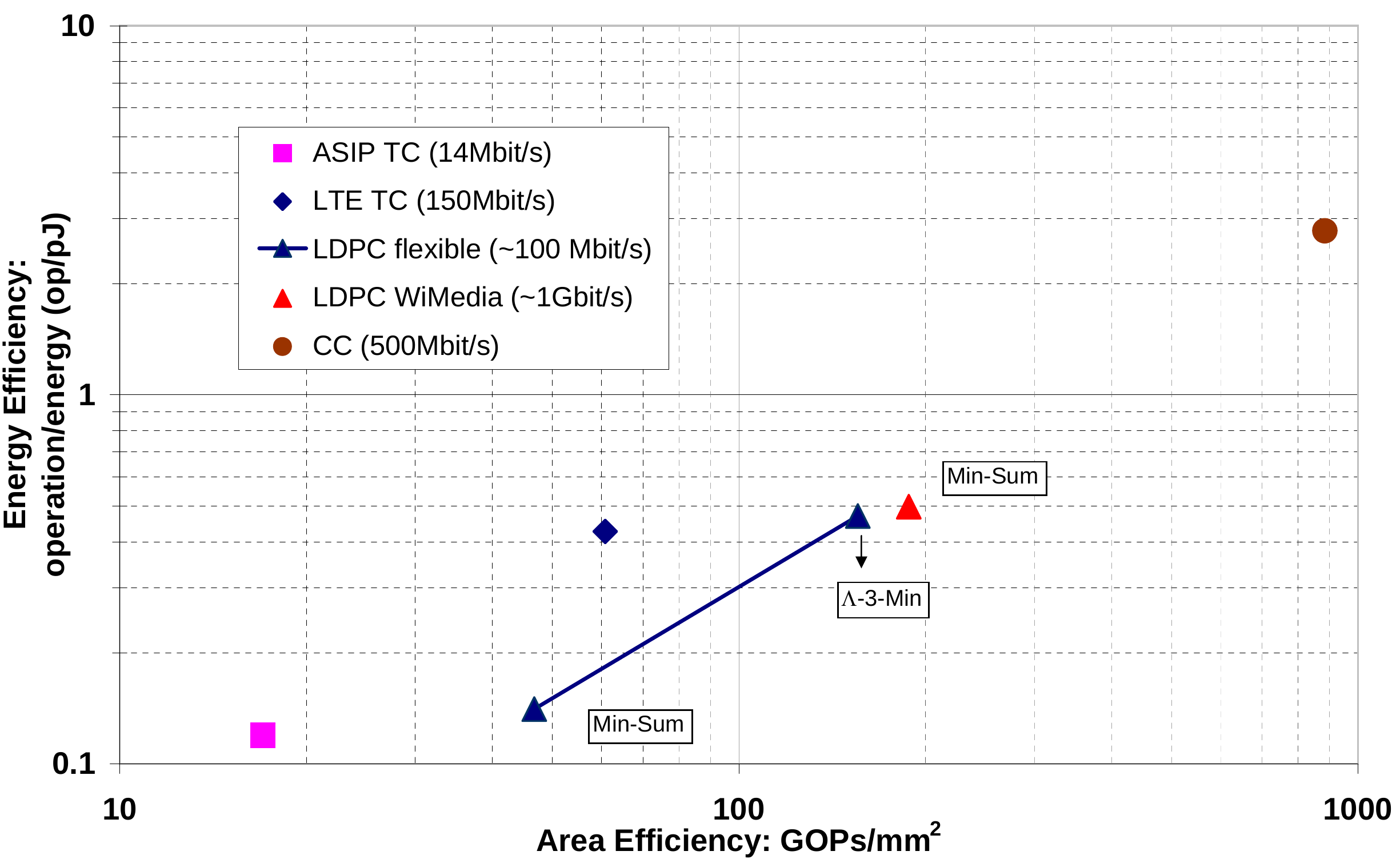}
\caption{Energy efficiency versus area efficiency based on operations}
\label{fig:EnergyOPS}
\end{figure}

The two dimensional design space, covering area and energy efficiency, is illustrated in \reffig{fig:EnergyOPS}. 
In this graph, efficient architectures w.r.t. area and energy have to be located in the upper right corner. 
Less efficient architectures are placed in the lower left corner.

We see that the convolutional decoder appears to be the most efficient decoder while the ASIP being the decoder with the lowest efficiency.
One interesting observation is the efficiency of the flexible LDPC decoder. The efficiency largely increases in both directions (area, energy) when 
replacing the Min-Sum by the $\lambda$-3-Min algorithms which is the more complex algorithm in terms of operations. 
As described in the previous section the GOPs for this algorithm increases by a factor 3. We
would expect a large increase in area and power accordingly. However, the area and power only increases by about 10\%
for the $\lambda$-3-Min algorithm. This is 
due to the fact that area and energy in both decoders are dominated by the data-transfer and storage scheme 
and the change in the operations of algorithm has only a small impact on it. In other words, the number of 
operations increases much larger than the implementation complexity. This is a hint that a GOPs based metric 
is not suited. Moreover we see that
the $\lambda$-3-Min based flexible decoder has nearly the same efficiency as the less flexible WiMedia decoder which
is optimized for throughput. We would expect that such a less flexible for throughput optimized decoder has a higher 
efficiency compared to the flexible one.

In the following we introduce metrics to resolve the afore mentioned anomalies.
Instead of using the operations which have to be carried out
for processing per task we normalize to the number of information bits per task. Metrics normalized to the number of information bits have the following properties. They allow comparing

\begin{itemize}
\item	competing architectures for a given algorithm since the efficiency metrics are independent of the specific operations and data types used to execute the algorithm. All implementation issues like data-transfer and storage are taken into account since the metrics is oblivious to how the task has been executed. 
\item	different coding schemes as a function of the communication parameters (modulation, signal to noise ratio, bandwidth).
\end{itemize}
 
In particular, iterative decoding algorithm can be compared in a meaningful way to non-iterative algorithm. Energy efficiency is a multidimensional problem. The performance of the physical layer of a communication system depends on the transmit energy via the SNR at the receiver, denoted \textit{communication energy}, and the processing energy in the receiver to retrieve the information, denoted \textit{computation energy}. There exists an interesting trade off between communication and computation  energy which has to be exploited in advanced, adaptive systems. For example, iterative algorithm operates at much lower SNR than convolutional codes. Decreasing the SNR, however, results in an exponentially increasing complexity and energy consumption due to the large number of iterations required for decoding.

We define the two suitable metrics for implementation efficiency as follows

\begin{itemize}
\item energy efficiency metric: \textit{decoded information bit per energy} measured in $bit/nJ$
\item area efficiency metric: \textit{information bit throughput per area unit} measured in $Mbit/s/mm^2$
\end{itemize}

\begin{figure}[t]
\centering
\includegraphics[width=1.0\columnwidth]{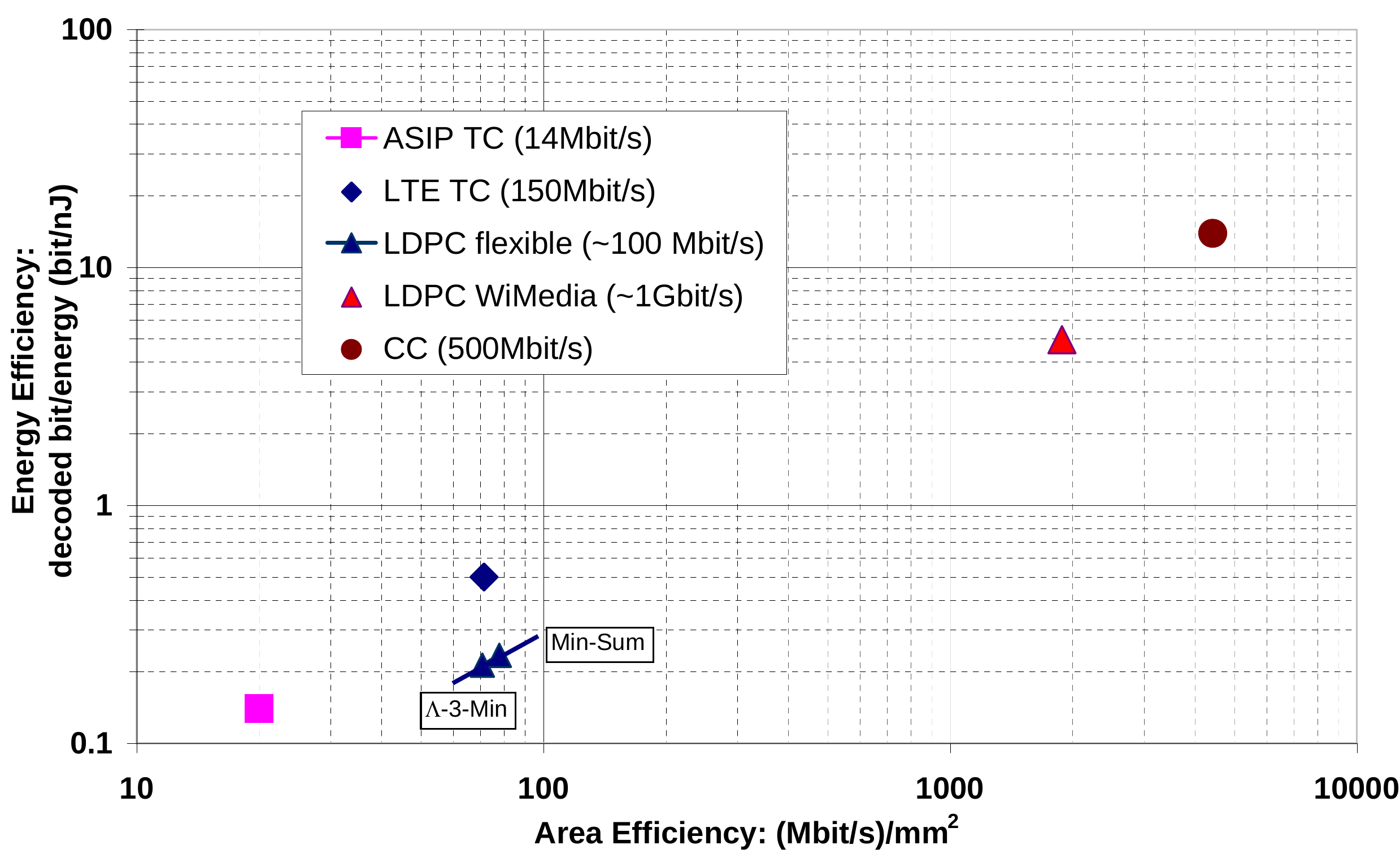}
\caption{Design space based on suitable metrics. Decoded information bit per energy over  information bit throughput per area unit.}
\label{fig:EnergyBit_metric}
\end{figure}

We have mapped our decoders to the design space which is based on these metrics, see \reffig{fig:EnergyBit_metric}.
Again efficient architectures are placed in the upper right corners, inefficient architectures in the lower left
corner.

A large change in the relative and absolute positions can be observed for some decoders, when comparing them 
with figure \reffig{fig:EnergyOPS}.

\begin{itemize}
	\item The difference in the efficiency between the two instances of the flexible LDPC decoder (Min-Sum and the $\lambda$-3-Min decoder respectively) is now much smaller. Moreover the Min-Sum decoder is more efficient than the other ones which was not the case in the conventional design space. This matches our expectations since the data-transfer and storage scheme in both decoders is nearly identical and the increase in computation results in only a small energy and area increase as described above. Both decoders are targeting the same throughput.
	\item The efficiency of the WiMedia decoder which is optimized for throughput and less flexibility, is now much
	      larger than the efficiency of the flexible LDPC decoder which again matches what we expected.
\end{itemize}

So far we have focused on the implementation complexity but have not discussed the important aspect of flexibility and communication performance. In the following we will investigate the relationship between communication performance, flexibility and implementation efficiency.

\section{Methodology}

In the previous section we investigated the absolute and the relative positions of the different decoders to each other. However equally important in this space is the trajectory when specific parameters are changed since such trajectory represents the impact of a specific parameter on the decoder efficiency. The following parameters will be considered: the frame error rates (FER), i.e. communications performance, coding techniques, code rates, number of iterations and throughput. The resulting trajectories well illustrates the strong dependency between communications performance and implementation efficiency.
We present two design space exploration methodologies. The first one is driven by implementation efficiency and compares non-iterative decoding techniques (convolutional codes) with iterative decoding techniques (LDPC codes). The second exploration compares two different iterative decoding techniques (LDPC and Turbo codes) with code rate flexibility.

\subsection{Implementation driven design space exploration}
\begin{figure*}[t]
\centering
\subfloat [Implementation Efficiency] {\includegraphics[width=3.5in, height=2.8in]{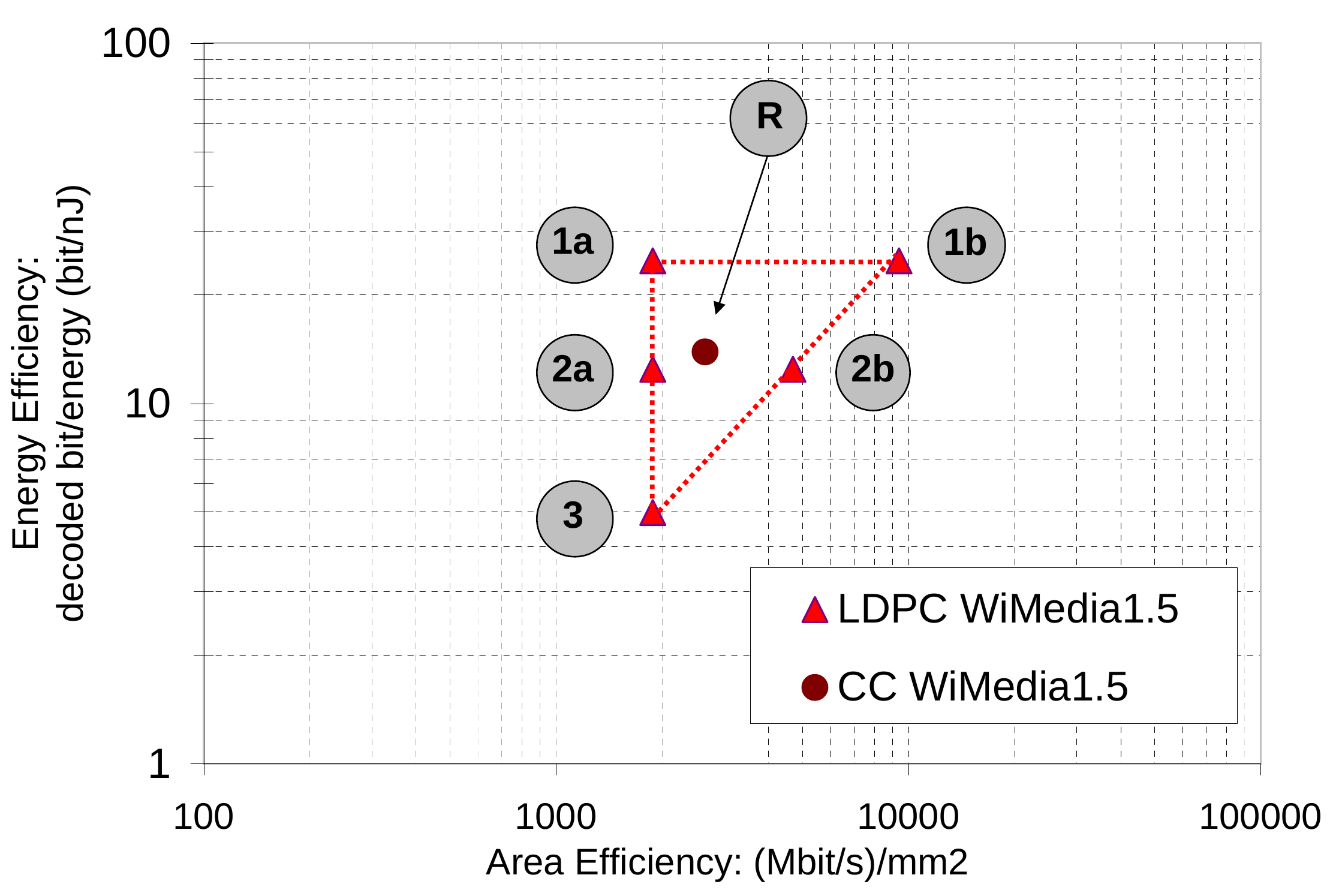}%
\label{fig:UWB_traj}}
\hfil
\subfloat[Communications performance (16-QAM, CM1 channel according to IEEE 802.15.3a \cite{allber_09}) ]{\includegraphics[width=3.5in, height=2.8in]{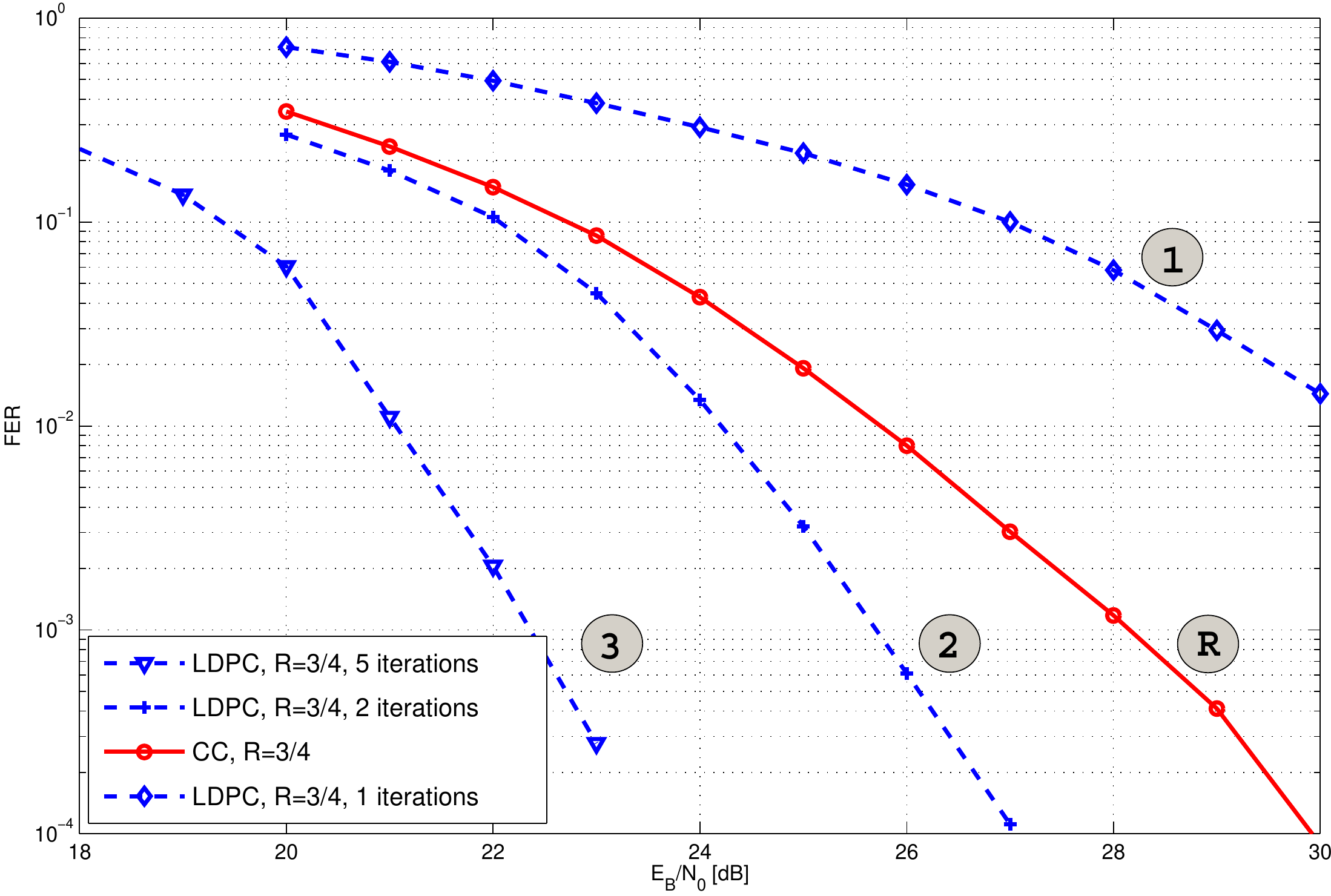}%
\label{fig:FER_UWB}}
\vfil
\fcolorbox{MyGray}{MyGray}{$
   \begin{array}{lll}
     \mathrm{R:} & \mathrm{Reference ~convolutional~decoder~with~code~rate~} (R=0.75)&\mathrm{ and~ fixed ~throughput ~of ~1~ Gbit/s } \\
     \mathrm{1:} & \mathrm{LDPC ~ code ~ performing ~ 1 ~ iteration} &  \\
     \mathrm{~~~~a)} & \mathrm{identical ~throughput ~(1 ~Gbit/s)} & \sim \mathrm{4dB~worse~communications~performance} \\
     \mathrm{~~~~b)} & \mathrm{5~times~ higher~ throughput~(5 ~Gbit/s)} & \sim \mathrm{4dB~worse~communications~performance} \\
     \mathrm{2:} & \mathrm{LDPC ~ code ~ performing ~ 2 ~ iteration} &  \\
     \mathrm{~~~~a)} & \mathrm{identical ~throughput~(1 ~Gbit/s)} & \sim \mathrm{1dB~better~communications~performance} \\
     \mathrm{~~~~b)} & \mathrm{2.5 ~times~ higher~ throughput~(2.5 ~Gbit/s)} & \sim \mathrm{1dB~better~communications~performance} \\
     \mathrm{3:} & \mathrm{LDPC ~ code ~ performing ~ 5 ~ iteration} &  \\
     \mathrm{~~~} & \mathrm{identical ~throughput~(1 ~Gbit/s)} & \sim \mathrm{4dB~better~communications~performance}     
   \end{array}
 $}
\caption{Implementation efficiency and communications performance for WiMedia 1.5 standard}
\vspace{-2mm}
\label{fig:ComparisonUWB}
\end{figure*}

We use the current WiMedia 1.5 standard for demonstration. WiMedia features low complexity devices for UWB communication. Thus the WiMedia 1.2 standard used convolutional codes as channel coding technique. In the definition phase of the next generation standard, WiMedia 1.5, LDPC codes were considered as a promising candidate due to their much better communication performance. A throughput of 960 Mbit/s at code rate $R=0.75$ was defined in the standard. A code/architecture co-design approach \cite{allber_09} resulted in an LDPC decoder which has a much higher efficiency than the flexible LDPC decoder. Note that its efficiency is lower in both dimensions compared to a convolutional decoder. However this comparison completely neglects the communications performance. Five decoding iterations can be maximally performed by the LDPC decoder to comply with the throughput fixed in the standard. As already pointed out, the number of iterations strongly impacts the performance of the LDPC decoder. 
The frame error rate as a function of the number of iteration is contrasted with implementation efficiency in \reffig{fig:ComparisonUWB}. Point 3 in the design space figure corresponds to the WiMedia 1.5 decoder when performing 5 iterations (this was the decoder assumption in the previous figures when we referred to the WiMedia LDPC decoder). The communication figure shows that this decoder has a 4dB better communication performance than the convolutional decoder. The communication performance is comparable to that of the convolutional decoder if the LDPC performs only two iterations instead of five (case 2 in \reffig{fig:ComparisonUWB}). Finally executing only one iteration in the LDPC decoder results in a communication performance which is about 4dB worse than the convolutional decoder (case 1 in \reffig{fig:ComparisonUWB}). 

Important is the resulting trajectory in the design space for the different cases. Two cases have to be distinguished: 
\begin{itemize}	
\item The system throughput is not changed w.r.t. WiMedia 1.5. constraint	(scenario $a$ in \reffig{fig:ComparisonUWB}). In this scenario only the energy efficiency 	is improved (points $3 \rightarrow 2a \rightarrow 1a$). Obviously the decoding time decreases when the decoder executes a smaller number of iterations resulting in a negative time lag. This time lag can be exploited for energy efficiency improvement. For example clock and the power supply could be completely switched off when decoding is finished. This reduces energy and leakage current. Another possibility is to slow down the frequency (frequency scaling). This reduces the energy by the same amount as in the previous case but the peak power consumption during decoding instead of leakage is minimized. The most efficient technique is voltage scaling in which	the voltage is reduced which results in the highest energy efficiency.   

\item The system throughput is changed (scenario $b$ in \reffig{fig:ComparisonUWB}). In this scenario the  area efficiency increases by the same amount as the throughput increases due to smaller number of iterations
(points $3 \rightarrow 2b \rightarrow 1b$).

\end{itemize}

We see that the efficiency of the LDPC decoder is increasing with decreasing communication requirements,
i.e. number of iterations. Thus, the \textit{decoder efficiency is represented by a trajectory instead of a single point} in the
design space. This trajectory results form varying communication performance requirements. We also see that 
the efficiency of the LDPC decoder outperforms the convolutional decoder at the same communication performance.

\subsection{Communications performance driven exploration}
\begin{figure*}[t]
\centering
\subfloat [Implementation Efficiency] {\includegraphics[width=3.5in, height=2.8in]{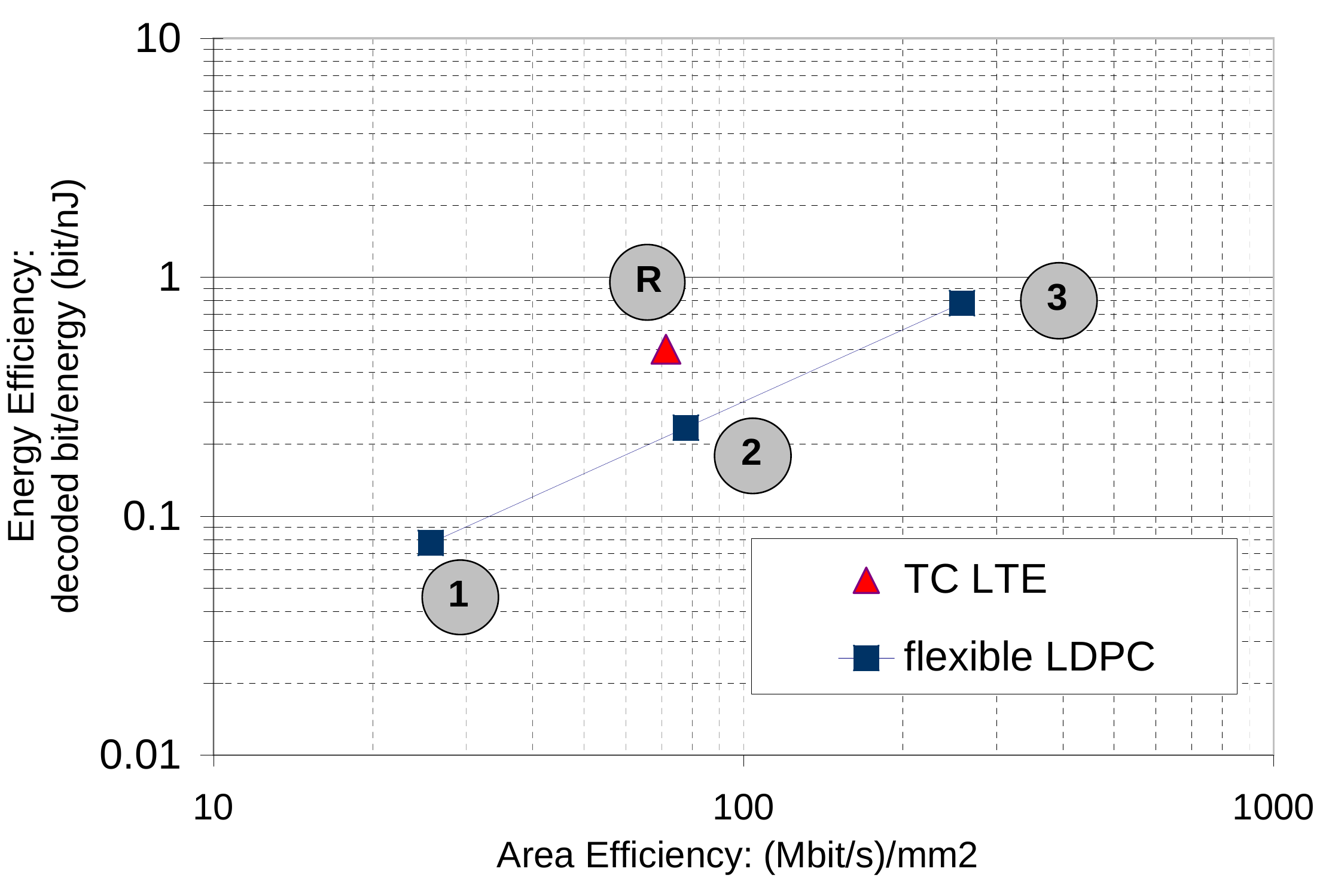}%
\label{fig:LTE_traj}}
\hfil
\subfloat[Communications performance (BPSK, AWGN channel)]{\includegraphics[width=3.5in, height=2.8in]{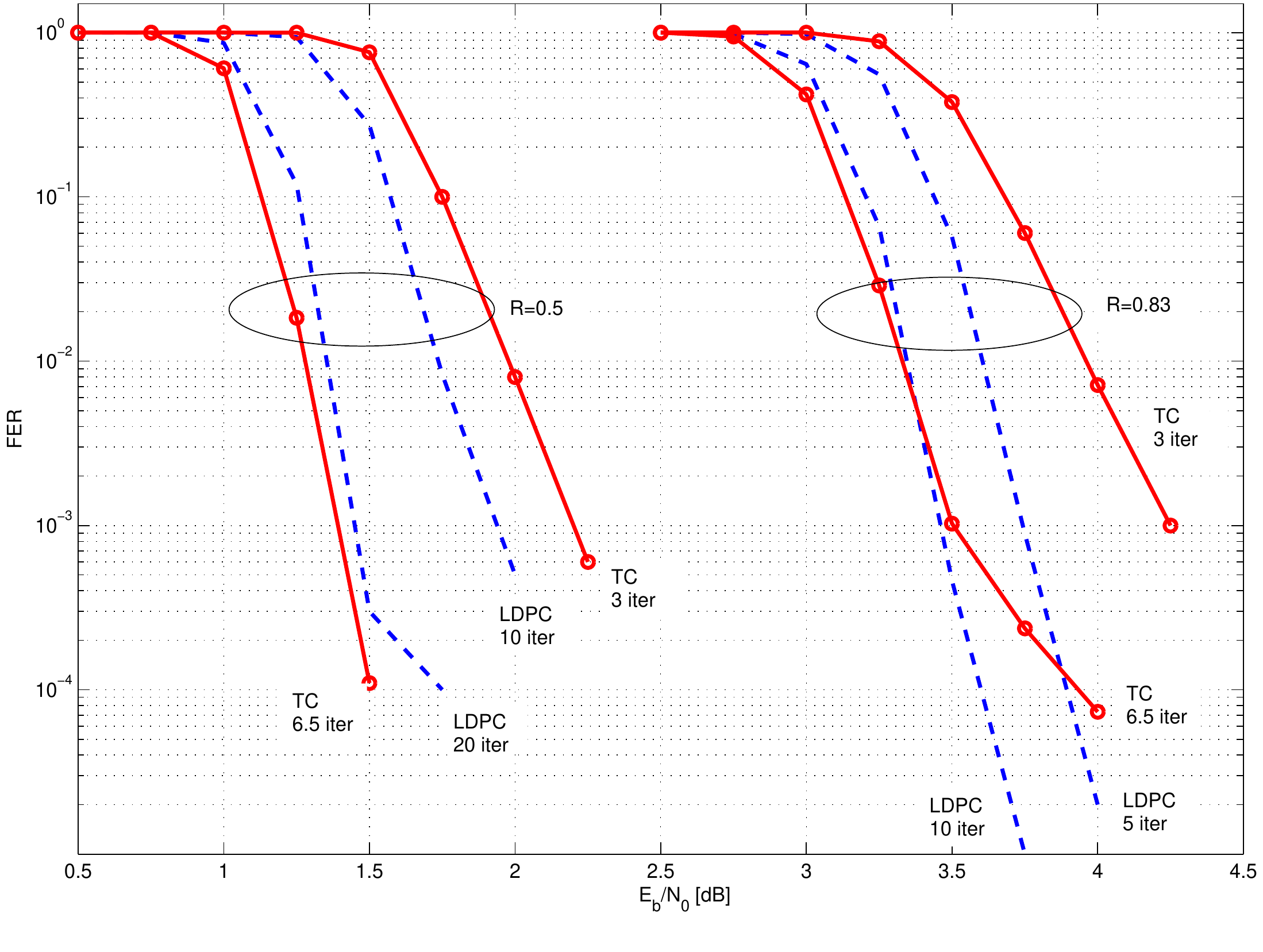}%
\label{fig:FER_LTE}}
\vfil
\vspace{2mm}
\fcolorbox{MyGray}{MyGray}{$
   \begin{array}{lll}
     \mathrm{R:} & \mathrm{Reference~of~LTE~turbo~decoder}&  \\
     \mathrm{~~~~} & \mathrm{150 ~Mbit/s ~throughput ~for ~all ~code ~rates~ } & \mathrm{reference ~TC ~performance ~at~ 6.5 ~iterations} \\
     \mathrm{1:} & \mathrm{LDPC ~ code ~ at ~ code ~rate ~} R=1/3&  \\
     & \sim \mathrm{ max ~throughput~30 Mbit/s} & \sim \mathrm{40~iterations~to~match ~TC~ performance} \\
     \mathrm{2:} & \mathrm{LDPC ~ code ~ at ~ code ~rate ~} R=1/2&  \\
     & \sim \mathrm{ max ~throughput~90 Mbit/s} & \sim \mathrm{20~iterations~to~match ~TC~ performance} \\
     \mathrm{3:} & \mathrm{LDPC ~ code ~ at ~ code ~rate ~} R=0.83&   \\
     \mathrm{~~~~} &  \sim \mathrm{ max ~throughput~300 Mbit/s} & \sim \mathrm{10~iterations~to~match ~TC~ performance} \\ 
   \end{array}
 $}
\caption{Implementation efficiency and communications performance of LTE turbo code/decoder and a flexible LDPC decoder.}
\vspace{-2mm}
\label{fig:ComparisonLTE}
\end{figure*}

In the previous exploration we have compared implementations efficiency between non-iterative and iterative decoding techniques dependent on
throughput and frame error rate behavior for \textit{fixed code rates}. In this section we compare two iterative decoding techniques and put
emphasis on code rate flexibility and dependencies. Reference is an LTE turbo decoder implementation. This LTE turbo decoder is compared with a flexible LDPC decoder which supports code rate flexibility. 
The right graph in \reffig{fig:ComparisonLTE} shows the communication performance for the
two decoding schemes dependent on code rates ($R=0.5$ and $R=0.83$) and iteration numbers. 
The number of information bits is $K=6140$ in all cases. Frame error rates are based on fixed point simulations matching the hardware
implementation.

We use the communications performance of the turbo decoder with 6.5 iterations as reference point for both code rates.
The 6.5 iterations result from the throughput constraint of 150Mbit/s which is specified in the LTE standard. The 6.5
iterations fulfill the LTE communications performance requirements for all code rates.

It is well known that the communication performance in LDPC 
decoding depends on the number of iterations \textit{and} the code rate. The LDPC decoder under investigations provide large
code rate flexibility, i.e., the hardware can support various code rates. 
The LDPC decoder requires 10 iterations for $R=0.83$ and 20 iterations for $R=0.5$ to match the performance of the turbo decoder. For a code rate of $R=1/3$ even 40 iterations are mandatory (this is not shown in \reffig{fig:ComparisonLTE}b).

Important are the corresponding trajectories in the implementation space. 
The turbo decoder efficiency is identical for all code rates (see left graph in \reffig{fig:ComparisonLTE}). Thus we have no trajectory. 
This is due to the fact that the code rate flexibility is implemented by puncturing which has negligible impact on throughput, area and energy.

However the situations is completely different for the flexible LDPC decoder. For a given communications performance the code rate has 
strong impact on the number of required iterations. This iteration number influences the implementations efficiency as we have seen in
the previous exploration case. But beside this impact via the iteration number, there is also a direct impact of the code rate on the 
implementation efficiency since lower code rates requires also a more accurate decoding algorithm ($\lambda$-3-Min algorithm instead of the
less complex Min-Sum algorithm). The resulting trajectory is shown in the left graph of \reffig{fig:ComparisonLTE}. We see that the 
efficiency increases in both directions with increasing code rate (points $1 \rightarrow 2 \rightarrow 3$).

The important observation in this exploration is the varying implementation efficiency of the flexible LDPC decoder represented
by the trajectory. This trajectory results from the required code rate flexibility in the LDPC decoder which is necessary to match
the communications performance with respect to a competitive turbo code decoder. We see that analyzing only one code rate, and thus 
one snap shot, could result in a wrong efficiency conclusions.

The two explorations have shown that implementation efficiency for advanced iterative decoders often results in a trajectory instead of 
a single point in the design space. These trajectories result from the strong interrelation between communication performance, flexibility
and implementation efficiency.

\section{Conclusion}

Understanding the trade-offs between implementation efficiency, communications performance and flexibility will be key 
for designing efficient baseband receivers.
Meaningful efficiency metrics are mandatory to explore and evaluate the resulting huge design space.
We introduced and discussed suitable energy and area efficiency metrics which are based on decoded information bit per energy and throughput per area unit. Various channel decoder implementations were utilized to examine these efficiency metrics with respect to the achieved communications performance and with respect to the decoder flexibility.
The presented methodology allows to systematically compare different realizations by jointly considering:
implementation efficiency, communications performance and flexibility.


\ifCLASSOPTIONcaptionsoff
  \newpage
\fi



%

\bibliography{Main_TCAS_ARXIV.bbl}

%

%
%
%




\end{document}